# A generic cloud migration process model

Mahdi Fahmideh[a], Farhad Daneshgar[b], Fethi Rabhi[c] and Ghassan Beydoun[d]

[a]University of New South Wales, Sydney, Australia; [b]SP Jain School of Global Management, Sydney, Australia; [c]University of New South Wales, Sydney, Australia; [d]University of Technology Sydney, Sydney, Australia

## Abstract

The cloud computing literature provides various ways to utilise cloud services, each with a different viewpoint, focus, and mostly using heterogeneous technical-centric terms. This hinders efficient and consistent knowledge flow across the community. Little, if any, research has aimed on developing an integrated process model which captures core domain concepts and ties them together to provide an overarching view of migrating legacy systems to cloud platforms that is customisable for a given context. We adopt design science research guidelines in which we use a metamodeling approach to develop a generic process model and then evaluate and refine the model through three case studies and domain expert reviews. This research benefits academics and practitioners alike by underpinning a substrate for constructing, standardising, maintaining, and sharing bespoke cloud migration models that can be applied to given cloud adoption scenarios.

## Introduction

Cloud computing technology has received significant attention in addressing the requirements of legacy software systems such as increasing computational power, reducing infrastructure costs, and the efficient utilisation of resources (Armbrust et al., 2010; Buyya, Yeo, & Venugopal, 2008; Koçak, Miranskyy, Alptekin, Bener, & Cialini, 2013). Many organisations are migrating their systems to cloud platforms and many others are moving from one existing cloud platform to another in a post-migration phase. The global cloud computing market continues to grow from $40.7 billion in 2011 to an expected $241 billion in 2020 (Ried, 2011).

Concomitant to this growth has been a volume of research conducted by both academia and industry. The research studies vary from the purely technical to the highly theoretical ones analysing organisational impact of the cloud technology (Oliveira, Thomas, & Espadanal, 2014; Venters and Whitley, 2012; Yang and Tate, 2012). Many new concepts have thus been bandied around the cloud computing domain. These concepts relate to various migration activities from planning, platform selection, reengineering, code refactoring, testing to legacy system deployment on cloud platforms. Depending on the research source, concepts are expressed using different terms, fragmented, or even merged in the literature.

As with any new emerging field, the interwoven tapestry of languages is initially profitable in allowing an interpretative and a creative discourse in the inception phase of the field. But, as the field of cloud computing somewhat matures a consensual view of the cumulative research that binds and integrates all the views together is more efficacious for knowledge sharing (Hollenbeck, 2008; Whetten, 1989). Developing that view requires a conceptual foundation which is not yet available. This gap hinders knowledge interoperability and critical information sharing among scholars and/or practitioners involved in cloud migration scenarios (M. Hamdaqa and Tahvildari, 2012; O. Zimmermann, Miksovic, & Küster, 2012). Making a separation between the definition of a cloud migration process and technical platform-specific operationalisation details is also becoming more pressing (Hamdaqa, 2011). Indeed as noted in Fahmideh et al. (2016), the time is ripe for providing a more abstract view of the current chaotic state of cloud migration.

There are many common concepts incorporated into existing models related to migrating systems to the cloud. Although they may not have been expressed in an identical way, it can be helpful if these concepts such as phases and activities are factored out into one, or at least be subsumed by one, unified model at a convenient level of abstraction. Each year a considerable number of models are suggested, each represents a different viewpoint of the same conceptualisation. This in itself is an indication that the field has reached a maturity point where the development of such a generic reference model becomes timely and important.

Metamodels capture common concepts, relationships, and ways of working. They are often suggested for achieving knowledge interoperability and integration of a domain of interest (Atkinson and Kuhne, 2003; Gonzalez-Perez and Henderson-Sellers, 2008). In essence, they provide a language infrastructure to freely describe the domain in a way that users can better understand it (Rossi & Brinkkemper, 1996). The significance of metamodels as a way for abstracting cloud computing concepts has been already emphasised in the community (M. Hamdaqa and Tahvildari, 2012; Leymann, 2011; Loutas, Kamateri, Bosi, & Tarabanis, 2011). In line with this view, the objective of this paper is to *develop and evaluate a generic metamodel that captures and harmonises common process elements of cloud migration process and that can be used to create, standardise, and share situation-specific cloud migration models.* Using the literature, we identified and distilled common concepts and integrated them into a process metamodel which is evaluated and refined using industry cloud migration exemplars. The resultant metamodel is cloud computing specific but context agnostic. It can be grounded and extended to be adapted to a given scenario and provides a basis for method engineers to define, configure, and share any migration methodological knowledge for managing cloud migration endeavours.

The paper is structured as follows: The next section reviews the prior literature on cloud migration and use of metamodels. The Research method section presents the adopted research approach undertaken to develop the metamodel. This is followed by the Demonstration section which illustrates the expressiveness of the metamodel in representing enacted cloud migration processes. Next, the evaluation of the metamodel is presented. Finally, the paper concludes with a discussion on the implications and limitations of this study.

# Background and related work

## Legacy systems and cloud migration

Legacy systems are often characterised by maintenance challenges. A notable early definition of them is given by Bennett (1995) as "large software systems that we don't know how to cope with but are vital to our organisation" (p.19). They are costly to maintain, inflexible to changes, difficult to be integrated with other systems, and have outdated documentation. Nevertheless, they are major components of IT-based organisations, providing business services, organisational knowledge, and a significant competitive advantage (Bennett, 1995; Erlikh, 2000; Sneed, 1995).

The migration phenomenon is about the physical movement of people, i.e. migrants, from one geographic location to another for a certain period of time (Clark, 1986). Migration can be taken for a short term or long term, short distance or long distance, voluntary or obligatory, and have some permanence, clear source, and target locations (Lee, 1966). Studying cloud migration as an instance of switching from on-premise hosted systems to cloud-hosted systems has been studied from various perspectives. Some studies centre on the development of models aiding organisations to decide the suitability of cloud adoption (Misra & Mondal 2011). Others provide tools aiding for making decisions about selecting suitable cloud services (Khajeh-Hosseini et al. 2012). Others highlight inhibitors and enablers of using cloud services (Oliveira et al. 2014), and benefits of cloud adoption in terms of enhanced competitive advantages (Truong & Dustdar, 2010), or service interoperability issues (Toosi et al., 2014).

Often a cloud migration process involves many concept variants and several ways of instantiation. This process itself is contingent on existing organizational structures and characteristics of a legacy system. It is common wisdom that no two cloud migration scenarios are exactly the same and every scenario requires its own specific management process. Failures in migration scenarios are often due to poor understanding of computing requirements, early engagement with the technical implementation of a cloud solution, and facing unexpected issues that were out of the control of service consumers and providers (Chow et al., 2009; Linthicum, 2012; Pepitone, 2011; Tsidulko, 2016).

A model at a conceptual level which aims at identifying core domain concepts and their relationships can help zoom in and provide a foundation for the representation and maintenance of bespoke cloud migration models. This not only assists method engineers in managing complexity but also allows sharing method knowledge among varying cloud computing communities. More recently, Fahmideh et al. (2016) have reviewed existing migration models and found that each one comes with its own concepts with a varying focus such as reusing legacy system codes (Menychtas et al., 2013), addressing interoperability issues (Mohagheghi et al., 2010), and finding optimum distributions of system components on cloud servers (Frey and Hasselbring, 2011; Menzel and Ranjan, 2012). Beyond this technical material, which is still important, we are yet to see a research effort that provides an integrated picture of the various methodological concepts. Tying those fragmented works within the literature and making an integrated view of intellectual bins is certainly compelling. Metamodelling is a clearly plausible approach as we shall discuss next.

## Metamodelling

A metamodel is "a model of a model or a model of a collection of models" (Atkinson & Kuhne, 2003). Raising the level of abstraction in modelling systems along with advantages such as improved reusability, interoperability, and reduced system development time has resulted in the emergence of a large number of metamodels. To provide a synopsis of notable literature on using metamodels to facilitate the use of cloud computing technology, we identified four streams as follows.

The first stream concentrates on abstracting the technical aspects of a cloud computing architecture such as multi-tenancy, elasticity, and data security. Studies such as (M. Hamdaqa, Livogiannis, T., Tahvildari, L., 2011; Liu et al., 2011; Zhang and Zhou, 2009; A. Zimmermann, Pretz, Zimmermann, Firesmith, & Petrov, 2013) and white papers published by major cloud computing players such as IBM, HP, Oracle, and Cisco are subsumed under this classification. Capturing the common knowledge of designing solution architectures has been the topic of discussions in (Fehling, Leymann, Rütschlin, & Schumm, 2012; Fehling and Retter, 2011) in which researchers propose a catalogue of software patterns for integrating legacy source codes with third party cloud services.

The second stream uses metamodels as a way for sharing green cloud computing practices such as reducing energy consumption and carbon emission of data centres (Procaccianti et al, 2014). Another work proposes a metamodel of the green practice for business processes leveraging cloud services (Nowak et al, 2014). These include classes of patterns for environmental impact, pollution, and waste. Dougherty et al. (2012) also propose a metamodel-based auto-scaling resource management to improve server utilization and reduce idle time compared with over-provisioned servers.

The third stream is concerned with quality aspects of cloud services. A metamodel developed by the A4Cloud project captures the knowledge related to non-functional properties of cloud services and how they influence the accountability of their providers (Nunez et al, 2013). The purported goal of the metamodel is to act as a language to model cloud service accountability in terms of transparency, verifiability, observability, and liability from which metrics are derived to monitor the quality of cloud services. The proposed metamodel by Cimato et al, (2013) models concepts related to the certification process of cloud services. Keller & König (2014) and Martens & Teuteberg (2011) respectively propose metamodels to model risks and compliance efforts for cloud computing as a socio-technical artefact.

The fourth stream uses model-driven techniques to represent legacy system variabilities combined with transformation to a given target cloud platform. For example, studies by (Ardagna et al., 2012; Wettinger et al., 2013) address the issue of migrating a legacy system across different cloud platforms using metamodel transformation techniques. Research in this direction has also resulted in languages in areas such as risk modelling (Zech et al, 2012), service compliance management (Brandic et al, 2010), cryptography (Bain et al, 2011), distributed data-parallel computing (Isard & Yu 2009), cloud-mobile hybrid applications (Ranabahu et al, 2011), big data analytic algorithms (Weimer et al, 2011), automatic code generation (Sledziewski et al., 2010), and maximizing reusability of SaaS (software as a service) (La & Kim, 2009). The central claim of these technical studies is on the seamless transformation of legacy system codes to different cloud platforms using model transformation techniques. The current research develops a metamodel that raises the abstraction level to cloud migration process.

# Research method

## Overview

We developed our metamodel using the design science research paradigm (Gregor and Hevner, 2013; Peffers, Tuunanen, Rothenberger, & Chatterjee, 2008). Design science research typically involves developing new artefacts, constructs, models, methods, or instantiations to address organisational IT problems. We conducted phases proposed by Peffers et al. (2008). As shown in Figure 1, the input to each phase was the metamodel resulted from the predecessor phase. We conducted the following phases in the one iteration.

**Problem identification and objective definition.** The proposed metamodel is addressing an important and timely problem with respect to a constituent community: different and heterogeneous viewpoints of the same process of legacy system migration to cloud platforms. Each viewpoint is expressed with different terms that are narrow in focus. There is currently no established mapping between these viewpoints to attain a harmonised overarching view. The proposed formative metamodel describing the process required for moving legacy systems to cloud platforms can be potentially a candidate for future method tailoring and interoperability in a consistent and systematic manner. It is agnostic to both the target cloud platforms and the domain of legacy systems.

**Design.** We identified a set of commonly used process concepts along with their definitions and relationships from the literature on cloud migration. The differences between definitions were reconciled into a consistent and coherent set of concepts. These concepts were grouped based on their similarities/context and then organised into a generic process metamodel. The outcome of this phase was the first version of the metamodel, version 1.0.

**Demonstration and evaluation.** We have used two methods for the validation. Firstly, the phase examined the expressiveness power of the first metamodel version 1.0 in representing process elements of three projects. The selection of the projects for the analysis was based on (i) having clear goals of cloud migration, (ii) adopting various service delivery models such as IaaS (infrastructure as a service), SaaS (software as a service), and PaaS (platform as a service), and (iii) availability of projects' databases including documentation, diagrams, notes, and codes related to the conducted scenario. The metamodel was found deficient with respect to some new concepts that were earlier overlooked. Support for these new concepts was added to the metamodel. This yielded the next version 1.1 of the metamodel.

Next, the metamodel was reviewed by a group of domain experts. Based on input from our interviewees and other informal communications with different practitioners and academics working in cloud computing, we noted that identifying expertise for the purpose of metamodel evaluation was somewhat difficult. Such expertise is often misconstrued and labelled in the midst of a myriad of other IT areas of expertise. This is attributed to two reasons: (i) the fact that much of legacy system migration to cloud projects are conducted partially due to barriers such as security issues and unwanted organisational changes, (ii) a subjective interpretation of the cloud migration process where some people viewed it as merely virtualization and others deemed it as huge legacy code refactoring. Considering every willing person to conduct metamodel evaluation was clearly not scientifically sound. Our general rule-of-thumb was the selection of experts who had hands-on experience and been directly involved throughout cloud migration projects as a programmer, system architect, or project manager for at least one year. Alternatively, an expert could also be an academic with scientific publications in peer-reviewed journals in the cloud computing field related to migrating legacy systems to cloud computing platforms. We carefully scanned profiles of cloud computing experts in the social media such as Twitter, Linkedin, Facebook, and academic research groups. We recruited only if the identified professional had experience in the roles above and/or high-level academic knowledge on legacy system migration to the cloud. For selected experts, the metamodel description (a twenty-five pages document detailing the metamodel along with a list of open-ended questions about the metamodel) and an invitation letter were sent to the expert in a subsequent communication email. We identified four experts from different countries who were interested in conducting the metamodel review. They all had leading cloud migration roles exceeding 7 years. The experts were not aware of and did not communicate each other. Their feedback was analysed and used to refine the metamodel. During the analysis, clarifications were sought as needed to prevent any misinterpretation of their comments. The output of this phase resulted in the final version of later metamodel presented in Figure 2.

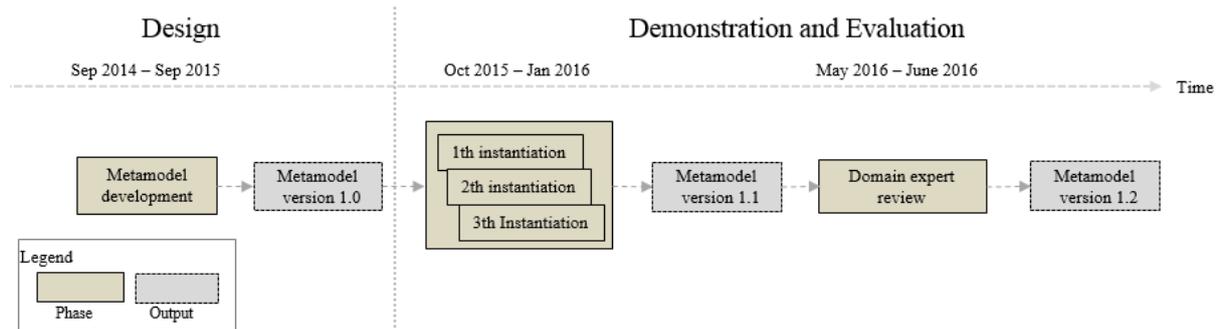

**Figure 1** Design science research process specialised for this research

## Design Phase

**Developing design principles for the metamodel**

Assuring the quality of a metamodel is an integrated part of the actual metamodelling process which bears on the metamodel ability to satisfy stated needs. Lindland et al. (1994) proposed general design principles (DPs) underlining a metamodelling endeavour. These are the following: syntactic adequacy, tailorability, and comprehensibility. The way each principle has been applied in the context of this research is detailed in what follows.

**Semantic adequacy (DP1)** is the correspondence between concepts in the metamodel and the domain of interest, i.e. cloud migration processes. To ensure this, two evaluation criteria are applied during the metamodelling process: *completeness* and *validity*. Completeness is the extent to which a metamodel can be used to make statements about the domain. Validity is the correctness of statements and their relevance to the domain. Achieving full adherence to this design principle may not be practical but an appropriate coverage of core process concepts incorporated during a typical transition is important. In this research, a good yardstick to get a feasible semantic quality is key functional and non-functional methodological requirements specific to cloud migration as elaborated in Fahmideh et al. (2016). These include, for example, analysing organisational context, identifying computational requirements, understanding legacy system architecture, and the choice of target cloud platform. Additionally, a metamodel needs specifying relationships among process components such as sequences, associations, and aggregations. For example, according to Fahmideh et al. (2016), a common challenge in migrating systems to the cloud is incompatibilities (e.g. data or functions) between legacy systems and cloud services. That is, for a chosen cloud platform, a sequence of steps in the migration process is required to identify any incompatibilities between platforms. Relationships defined in our proposed metamodel are based on recommendations in the literature. Our overall definition of semantic adequacy is: *the proposed metamodel should capture all important relevant concepts for the incorporation into a typical transition process of legacy systems to cloud platforms*.

**Tailorability (DP2)** is the extent to which a metamodel can be customised and extended to address new requirements. A process metamodel allows tailoring into different models. Tailorability is required as the integrating legacy systems with cloud services may be undergone by several factors such as the choice of a target cloud platform, reusability of legacy system codes, security requirements, and system workload. These factors and many others influence tailoring decisions. To support tailorability, the metamodel should have some modularity in a way that different subsets of concepts can be selected and put together to fit needs of a particular scenario Cameron (2002). Another prerequisite for the tailorability is the fact that the more a metamodel is close to the problem domain, the simpler is its customisation and specialisation (Jonkers et al, 2006). Therefore, another key attribute of an effective process metamodel is: *the metamodel should be tailorable to different cloud migration scenarios.*

**Pragmatic quality (DP3)** is the extent to which a metamodel is comprehended by its audience (Lindland et al, 1994). A metamodel is expected to have understandable concepts, to reflect intentions of its audience, to minimise multiple interpretations, and to avoid unnecessary modelling details (Ambler, 2005). This quality is largely determined by the quality of its diagrams, icons, names and definitions (Lindland et al, 1994). In the context of this research, the third design principle is formulated as follows: *the definitions, names, and relationships of concepts in the metamodel should be comprehensible by cloud computing domain experts.*

Metamodels have the potential to become too complex if they include a large number of concepts, definitions, and relationships. In deciding on the level of metamodel complexity, every designer faces a trade-off between tailorability, understandability, and comprehensiveness Gonzalez-Perez & Henderson-Sellers (2008). If the designer tends to maximise the metamodel tailorability, then abstracting out and making domain concepts variable instead of being fixed takes precedence over understandability. On the other hand, adherence to the understandability principle pushes towards making the metamodel more detailed and elaborated at the expense of the tailorability. Making a trade-off among these principles, e.g. too generic or too specific metamodel, is always a difficult issue to decide (Henderson-Sellers & Gonzalez-Perez 2010). For example, bearing in mind the priority of the metamodel understandability over the completeness, the designer may not include many different complex domain concepts and relationships in the metamodel to make a more generic and less detailed metamodel. In the current study, results identified from examining the metamodel adherence to its purported design principles through conducting the case study analysis and domain expert review (sections Demonstration and Evaluation) show the metamodel is not a complicated entity to use.

**Metamodel development**
A brief explanation of steps undertaken to develop our initial metamodel are explained in what follows, though a more detailed description is available at Fahmideh et al. (2017b). To create the metamodel, we conducted a systematic literature review (Kitchenham et al, 2009) as a point of departure for identifying cloud migration process concepts. These concepts could be a (i) *task*: a discrete and small unit of work performed by developers to achieve specified goals, (ii) *work-product*: a significant artefact as a result of performing tasks, (iii) *principle*: a consideration that should be taken into account during design of a cloud solution architecture, and (iv) *phase*: a collection of concepts logically classified to provide a high-level organisation to the cloud migration process.

Adherence to the design principles (DPs) has been interleaved with the actual metamodelling process to ensure that the metamodel satisfies stated needs. To address DP1, we had a tendency in selecting concepts that were cloud-platform independent and sufficiently generic to a variety of cloud migration scenarios. Concepts that were too general or belonged to the traditional software development process were not incorporated as they were deemed out of the scope of this research. The full list of all the identified studies along with the extracted concepts is presented in the Appendix A. For DP2, it was critical in the metamodeling process that concepts are chosen at the right abstraction layer. Through a bottom-up approach, all concepts were grouped based on their similarities and definitions to derive a new set of high-level overarching concepts. Various definitions of concepts were reconciled to reach a set of internally consistent metamodel concepts. When there were several definitions for a concept, a hybrid definition encompassing all definitions was synthesised. The relationships among all process concepts such as sequence, association, specialization, and aggregation were revisited as needed. For DP3, the choice of concepts' names, definitions, and terms was made in a way to be intuitively understandable for users. A simple version of Unified modelling language (UML) notation UML (2004), which is common for information modelling, was used to represent the metamodel in a well-structured manner.

# Resultant Metamodel

In the following, we provide an overview of the metamodel though a more detailed technical description of its internal working can be found at Fahmideh et al. (2017a). The metamodel includes a set of concepts that are commonly performed in the cloud migration. They are organised into three phases namely *Plan*, *Design*, and *Enable*. Operationalisation details of concepts are left to each individual instantiation of the metamodel using available implementation techniques in the cloud computing literature or tools in marketplace. Figure 2 shows the metamodel along with definitions of the concepts presented in Table 1.

The *plan* phase starts with a feasibility analysis of adopting cloud services in terms of potential changes in organisational structure, local network, and cost saving. The legacy system architecture and its functional and non-functional requirements are identified. This can be a deployment model of the system in the local network of organisation. This model helps in estimating required effort to make the system cloud-enabled. Legacy systems may have certain requirements that can be satisfied by utilising cloud services such as computational, storage space, or security requirements. The phase also includes preparing a plan which organises the sequence of activities in the course of a migration process.

In *design* phase a new architectural model showing how legacy system components utilise cloud services is produced. The re-architecting process includes identifying suitable components for moving to and re-deployment in the cloud in order to satisfy non-functional requirements such as data security, performance variability, acceptable network delay, and scaling latency. In re-architecting legacy systems to the cloud, design principles play a central role. For instance, system components should minimise dependency and store the contextual data during their execution in order to support the individual scalability feature. An important consideration during cloud architecture design is the performance variability of cloud servers and latency between a local network and cloud servers. Developers should implement mechanisms in legacy systems to detect and handle transient faults that may occur in the cloud. A key work-product of this phase is a new skeletal architecture specifying an optimum distribution of legacy components on the cloud servers with respect to non-functional requirements.

The *enable* phase is to implement the architectural model designed in the previous phase. Legacy systems might have been implemented with technologies that are not compatible with cloud services. If this situation occurs, such incompatibilities should be identified and accordingly resolved through adaptation mechanisms such as refactoring source codes of legacy systems, modifying data, and wrappers. Legacy systems might not have been implemented with the support for dynamic resource acquisition and release under input workload. Instead, new physical servers are added to address workload. Mechanisms for system elasticity are implemented by continuous system monitoring and performing actions for resource management based on scaling rules triggered in a specific workload threshold, event, or metric. The phase may entail either adding new components to the new legacy system architecture or having them separately hosted on cloud servers. Additionally, the local network is reconfigured to provide access to cloud services. If required, legacy components and third party tools are installed. Finally, both functional and non-functional aspects of the migrated system are tested.

**Figure 2** The proposed metamodel

Table 1. Key process concepts incorporated into a typical process of legacy system migration to the cloud

| Concept | Definition |
|---|---|
| Analyse context | Analyse migration suitability with respect to factors such as cost of legacy system modification, installation, training, administration, license management, required expertise, pricing models of the service providers, infrastructure procurement imposed by the migration, impact of the cloud on stakeholders, organisational constraints, responsibilities, and working practices. |
| Analyse migration requirements | Identify a set of requirements to be satisfied by the cloud such as computational requirements, data storage, security, response time, and elasticity. |
| Define plan | Define a sequence of tasks that guide the migration process by analysing feedback from stakeholders. A plan may include (i) notice of temporal unavailability of legacy systems, (ii) roll-back the system to in-house versions, (iii) migration type such as complete or partial, and (iv) legacy system retirement procedures. |
| Recover legacy system knowledge | Produce a complete representation of legacy system architecture including its data, components, dependencies among components and infrastructure, system data usage, and resource utilisation model (e.g. CPU, Network, and storage). |
| Choose cloud platform/provider | Define a set of suitability criteria that characterise desirable features of cloud providers including pricing model, constraints, offered QoS, electricity costs, power and cooling costs, organisation migration characteristics (migration goals, available budget), and system requirements. |
| Design cloud solution | Identify legacy system components with respect to migration requirements and then define their distribution cloud servers. |
| Identify incompatibilities | Identify incompatibilities between legacy system components and cloud services. |
| Make system stateless | Enable the legacy system to handle safety and traceability of tenant's session when various system instances hosted in the cloud. |
| Decouple system components | Decouple system components from each other. Use mediator and synchronisation mechanisms to manage interaction between the loosely coupled components. |
| Replicate system components | Partition and deploy legacy system components (e.g. database, business logic) on multiple cloud servers. |
| Make mock migration | Build a prototype of new cloud solution to get an understanding of how the functional and none-functional aspects of the system will work in the cloud. |
| Use logging | Use logging mechanism to facilitate system debug and resource monitoring when running in the cloud. |
| Resolve licensing issues | Define and monitor a pay-as-you-go licensing model to handle unintended license agreement violations due to automatic scaling. |
| Develop integrators | Develop mediators/wrappers to hide incompatibilities occurring at runtime between legacy system components and selected cloud services that are plugged to these system components. |
| Deploy system component | Install system components and any required third party tools in the cloud. |
| Enable elasticity | Define scaling rules and provide support for dynamic acquisition and release of cloud resources. |
| Encrypt database | Encrypt critical databases prior to hosting in the cloud. |
| Handle transient faults | Detect and handle transient faults may occur in the cloud. |
| Isolate tenant | Protect tenants' data, performance, and faults from other tenants, which are running on the same cloud server. |
| Encrypt/Decrypt messages | Secure messages transmission between the local components and those hosted in the cloud or distributed across multiple clouds using an encryption mechanism. |
| Obfuscate codes | Protect unauthorised access to code blocks of components by other tenants that are running on the same cloud provider. |
| Re-configure network | Re-configure the running environment of the system including reachability policies to resources and network, connection to storages, setting ports and firewalls, and load balancer. |
| Synchronise/ replicate system components | Provide support in the system to synchronise multiple components (e.g. database replica) hosted on premise network and cloud servers. |
| Communicate a-synchronous | Enable application components to interact in an asynchronous manner. |
| Test system | Test system security, interoperability, multi-tenancy, performance, scalability, network connectivity of the system that migrated to the cloud. |

# Demonstration and evaluation: retrospective case studies

As the proposed metamodel is sufficiently generic and covers key domain concepts, it is anticipated that real-world migration processes to be representable using the metamodel concepts. To appraise the metamodel adherence to DP1 and DP2, we used three case studies presented in Table 2. We traced the conformance of the concepts in these scenarios to the corresponding ones in the metamodel. This was performed by grouping and mapping concepts in these scenarios to the metamodel concepts according to their relevance. Some of the leading questions (see Appendix B) that were used during the case review were as follow: (i) what activities were performed and deliverables produced by the developers during each phase of the migration project? (ii) what cloud-specific challenges were that those developers faced in each phase?

Table 2 Description of case studies

| Case 1: *InformaIT* (Sweden) | Case 2: *TOAS* (Finland) | Case 3: *SpringTrader* (US) |
|---|---|---|
| *InformaIT* is a software development company providing digital document processing services. The Document Comparison (DC) system, developed by the company, is a Web-based enterprise solution for enhancing document management processes. DC provided a fast and easy way to compare textual and graphical content of different digital documents. DC was originally designed to offer services to medium and large organisations which had adequate infrastructure and staff to install and run the system. *InformaIT* wondered expanding DC's services around small publishing companies. However, small companies could not afford DC as they would need a high financial commitment for installation and paying and usage cost of users. Cloud-enablement of the system could facilitate an efficient and agile maintenance environment for the DC for small companies. | TietoOyj is a software development company that has built an open source platform called Tieto Open Application Suite (TOAS) that provides an integrated set of middle-wares, tools, and services for developing new software systems and deploying on the cloud. The platform aims increasing development speed, automation, and the integrity of cloud-based software systems. A cloud migration project was launched by Tieto to migrate a legacy batch processing system to this TOAS platform. The outdated hardware infrastructure and software platform of the legacy system was the key driver to move to TOAS which leads enhanced system performance and reduced infrastructure cost. | SpringTrader is an open-source Web-based system that has been originally developed by Pivotal company and maintained by many contributors over time. The system allows registered users to monitor and manage a portfolio of stocks, lookup stock quotes, and buy/sell stock shares. Pivotal company recently has developed its private cloud platform, which named Pivotal Cloud Foundry. The platform is an open-source platform for developing and deploying portable cloud-native enterprise systems. Pivotal decided to move SpringTrader system to this new cloud platform to enable users to access real-time stock market data in more interactive way, to individually scale up/down each system component (also called micro services), and to improve system maintainability. |

Inspired by previous studies suggesting the worthiness of secondary data in the assessment of metamodels (Antkiewicz, Czarnecki, & Stephan, 2009; Beydoun, et al., 2009; Othman and Beydoun, 2013), we used project documents from a variety of sources (e.g. system models, codes, and user histories) to obtain a better understanding of the enacted in-house method. The summarised results of the tracing in Table 3 show the extent to which metamodel adheres to DP1 and DP2. In this table, the first column shows a metamodel concept and the next three columns show corresponding instantiation of the concept in three case study scenarios.

## Within case analysis: InformIT case

The following paragraphs describe how the concepts in the metamodel are instantiated and specialised to represent tasks that were carried out by the development team in *InformaIT* project (Rabetski, 2012). The unit of analysis is the legacy system. The 43-page secondary document of this project was carefully reviewed. Figure 3 depicts instances of enacted metamodel concepts in *InformIT* highlighted with grey colour.

As one of the first tasks, the developers performed *preliminary analysis* to identify benefits and risk of migrating the system to the cloud in terms of privacy, vendor lock-in, and environmental limitations. This activity is an instantiation of the concept *analyse migration feasibility* in the metamodel. Additionally, an activity, called *current DC implementation*, was performed to identify the current deployment model of DC. The metamodel supports this task through an instance of *recover legacy system knowledge* defined in the plan phase.

The developers estimated the cost of the migration project on the basis of required server instances, storage, data transfer, storage transaction, cache, and database. They realised that the cost could be down from $764.99 in the cloud model compared to $1264.99 in the legacy system model when leveraging elastic scalability. The abovementioned cost analysis in *InformaIT* can be derived from the *analyse migration cost* in the metamodel which is a subclass of *analyse context*.

Once the cloud migration was perceived as a viable solution to empower DC, the developers performed an activity named *choosing a cloud provider* in order to analyse three candidate public cloud platforms Amazon Web Services, Google App Engine, and Microsoft Azure. Each candidate platform could affect the cost, the quality of the solution architecture, and the required legacy code changes. The developers found that Google App Engine could not be a suitable candidate for DC since it did not support .NET software systems unlike Amazon AWS and Microsoft Azure that both provided such a support. After a further analysis, the developers preferred Windows Azure platform for three reasons: (i) it would require less configuration effort, (ii) it would offer faster deployment model, and (iii) developers had a consistent experience in adopting Microsoft family technologies. The concept *Choosing a cloud provider* in *InformaIT* conforms to the concept *choose cloud platform/provider* in the design phase of the metamodel.

In *InformaIT* scenario the developers performed a task called *cloud DC architecture* indicating how the existing legacy components are mapped to the Microsoft Azure platform. For example, the legacy version of DC's database, a Microsoft SQL Server database, was replaced with SQL Azure. The metamodel generates this concept through an instantiation of concept *design cloud solution* in design phase of the metamodel.

Furthermore, the developers identified some incompatibilities between the legacy system and the cloud platform that implicated some changes to the current system implementation. This is referred to as *identified compatibility issues* in the metamodel. Subsequently, the migration process proceeded with some changes in DC. As an example, the data and queue storage technologies in Microsoft Azure were not compatible with regular application programming interfaces that were currently used by DC. Also, DC had been developed using Microsoft .Net 2.0 technology that was not supported by Microsoft Azure. The action to resolve this was to update DC's framework to Microsoft .Net 3.5/4. Other incompatibility issues were session management and registration of legacy components in the cloud. The classes *identify incompatibilities* and *refactor codes* in the metamodel represent the above modifications to the DC in *InformaIT* case.

We found that some changes to the DC were in the form of applying design principles defined by the concept *apply design principles* in the metamodel. For instance, DC was required to be portable between the local network and the cloud. To address this, the developers separated the data and business layers by adding a new intermediate data access layer. Hereafter, the business logic layer calls operations of the intermediate layer instead of a direct access to the data layer. In *InformaIT* project, this concept is called *separate data layer from business logic layer* which can be derived from the concept *decouple system components* as a subclass of the *apply design principles* in the metamodel. Moreover, DC stored megabytes of data per session that was a big overhead. Such a session size required more time for serialisation/de-serialisation. Developers applied a principle called *becoming as stateless as possible* in DC architecture. This concept is an instance of the principle *make system stateless* in the metamodel.

It was likely that the performance of DC in the cloud would be decreased due to unexpected latencies occurring in cloud servers. Developers used a small Azure compute instance and well as local server instance to perform *performance experiment* to execute CPU heavy code for the document rendering.

This enabled developers to compare the execution and response time of running DC in the cloud. The result of the experiment revealed potential performance bottlenecks in the cloud. The abovementioned test in this scenario conforms to the concept *test performance* in the enable phase of the metamodel.

Additionally, the suitability of the DC migration to the cloud was analysed from a cost perspective. Developers built a prototype to analyse three real life scenarios that could describe how DC could benefit from cloud services. The cost of each migration scenario was estimated based on the pricing model of Microsoft Azure, computing instance, relational database, storage transaction, data transfer, and cache size. Building a system prototype helped developers to make a final decision regarding the cloud enablement. The concept of the prototype in this scenario is representable by *make prototype* in the metamodel. Regarding DP1, the analysing *InformaIT* confirmed some relationships between the concepts defined in the metamodel. Table 3 shows the list of relationships among the metamodel's concepts that were instantiated in this migration scenario.

Within case analysis confirmed that majority of accommodated tasks in this scenario are derivable from the metamodel, except for a new concept *use logging* that was not covered by any concepts in the metamodel. The metamodel had a deficiency to support this concept. Since cloud environments are asynchronous, debugging and tracing a system in the cloud might be problematic (Rabetski, 2012). Applying a logging mechanism in the system architecture facilitates tracing system behaviour, resource utilisation, and identifying reasons for failures in the cloud. Therefore, the metamodel concept *apply design principles* was refined by adding a new concept named *use logging*. In Figure 2, this new concept is defined as the sub-class of *apply design principles*. It is defined as "Use the logging mechanism to facilitate the system debug and resource monitoring when running in the cloud". The inclusion of this new concept evolved the metamodel to the second version 1.1.

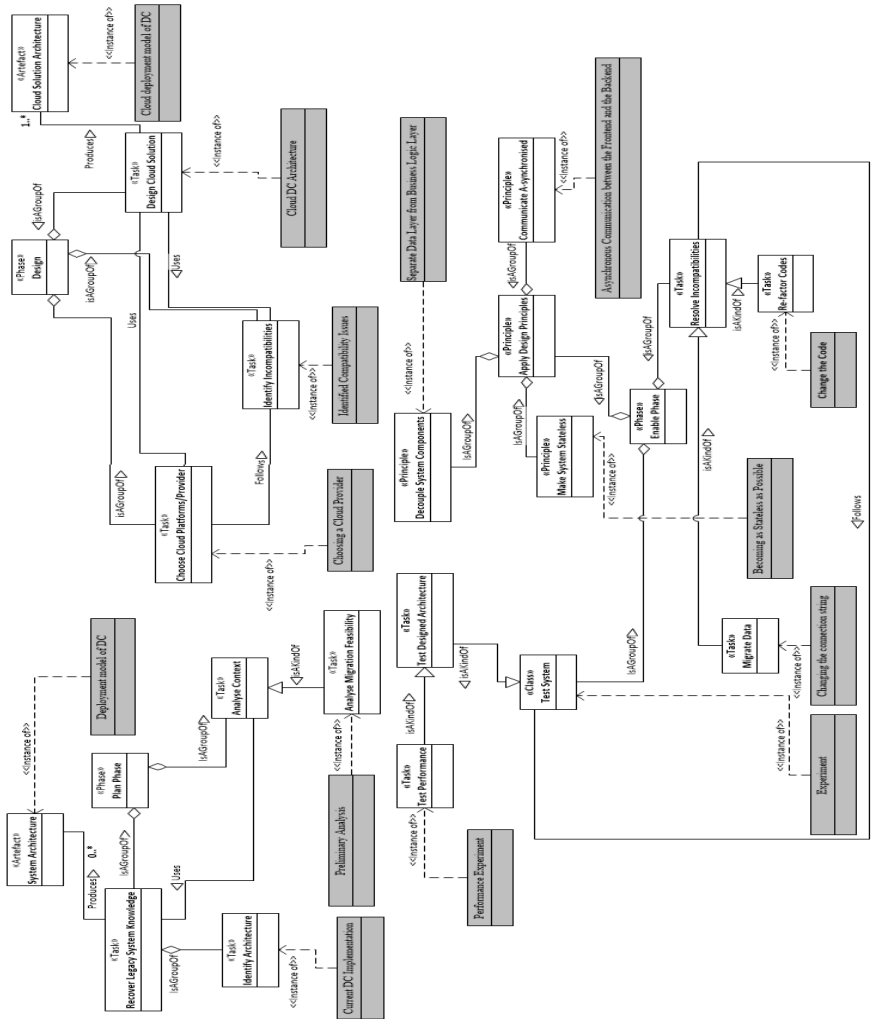
**Figure 3** *InformIT* model as an instantiation of the metamodel

## Cross case analysis

Our cross-case analysis examines to the extent the metamodel adheres to the DP1 and DP2 in each scenario. Table 3 shows the collection of the process concepts and relationships that were incorporated into the scenarios.

As for DP1, the review of the three scenarios shows that they instantiated four common metamodel concepts *recover legacy system knowledge*, *design cloud solution*, *identify incompatibilities,* and *decouple software components* in their mainstream process (see Table 3). For example, the concept *design cloud solution* in the metamodel was instantiated in three different ways in each scenario. In *InformIT*, the decision on the selection and deployment of legacy system components on cloud servers was basically a mapping between Microsoft-based legacy components and their counterparts in Microsoft Azure cloud platform. In TOAS, the legacy components were classified into two logical groups of platforms on the basis of similar functional behaviours. For *SpringTrader* case, components providing finance services were those selected for the migration purpose. These are an instance of *design cloud solution* defined in the metamodel.

Migration scenarios were conducted differently and therefore each scenario instantiated a slice of the metamodel elements to address its requirements. Except for *InformIT*, in both *TOAS* and *SpringTrader* scenarios the activities related to handling incompatibility issues were performed (seventh row in Table 3). In *TOAS* case, developers implemented run-time adaptors to resolve incompatibilities of message formats and interfaces between the legacy system and TOAS cloud platform. Comparably, in *SpringTrader* case, developers implemented wrappers to separate incompatibilities between cloud microservices and the legacy system. These techniques are subsumed under the concept *develop integrators* defined in the *enable* phase of the metamodel. Furthermore, unlike the instantiation of the concept *choose cloud provider* in the *InformIT*, where developers decided to use Windows Azure cloud platform due to their former experience in using this platform, the target cloud platform in both scenarios of *TOAS* and *SpringTrader* was a pre-chosen private cloud platform. Therefore, there was no need for the instantiation of the concept *choose cloud provider* (second row in Table 3). Moreover, the case studies also confirmed the correctness of some relationships among the concepts defined in the metamodel. Finally, the second and third case studies did not result in new refinements to the metamodel.

Table 3 Support of concepts and relationships in the migration scenarios by the metamodel (√:instantiated ×:not instantiated)

| | Name | InformaIT | TOAS | SpringTrader |
|---|---|---|---|---|
| **Metamodel concepts** | Recover legacy system knowledge | √ | √ | √ |
| | Choose cloud platform/provider | √ | × | × |
| | Design cloud solution | √ | √ | √ |
| | Identify incompatibilities | √ | √ | √ |
| | Decouple system components | √ | √ | √ |
| | Adapt data | √ | × | √ |
| | Develop integrators | × | √ | √ |
| | Refactor codes | × | × | √ |
| | Re-configure network | × | √ | × |
| | **Relationship** | **InformaIT** | **TOAS** | **SpringTrader** |
| **Metamodel relationships** | Design cloud solution *Uses* Analyse migration requirements | × | √ | × |
| | Design cloud solution *Uses* Identify incompatibilities | √ | √ | √ |
| | Design cloud solution *Uses* Choose cloud platform/provider | √ | √ | √ |
| | Refactor codes *Uses* Identify incompatibilities | √ | - | √ |
| | Design cloud solution *Uses* Recover legacy system knowledge | × | × | √ |
| | Refactor codes *Uses* design cloud solution | √ | × | √ |
| | Migrate database *Uses* Refactor codes | √ | × | × |
| | Test system *Uses* Design cloud solution | √ | √ | √ |
| | Plan migration *Follows* Design phase | √ | √ | √ |
| | Design phase *Follows* Enable phase | √ | √ | √ |
| | Choose cloud provider *Follows* Identify incompatibilities | √ | √ | √ |

# Evaluation and demonstration: expert reviews

The second version of the metamodel was qualitatively examined by a panel of four domain experts with respect to all design principles. The experts are denoted by E1, E2, E3, and E4. The questionnaire form to evaluate the metamodel is presented in Appendix C. The usefulness of the metamodel was stated by the words such as "education and high-level guidance" (E1), "good communication vehicle" and "more comprehensive list of concerns" (E2). E2 stated that "the model is clearly valuable in conveying the important concerns of a migration and how they are related. The detailed semantics help to clearly understand dependencies and possibly resulting decisions and trade-offs to be considered". A similar opinion was expressed by E3 who said "this model can make a good impact to increase the confidence of success factor of the migration process and decrease some uncertainty. Also, this model can be used as a checklist of successful migration and this reference model makes an overall picture of migration phase and clears the roadmap for audiences to do the migration with less stress and concerns". The advantage of the metamodel against existing models was stated by E4: "I have mostly used the classical reengineering model for legacy migration. In comparison to the model by SEI, the proposed model is more detailed in terms of underlying process and activities for migration". In view of the design principles, here are some suggestions for the improvement of the metamodel and refinements as a consequence of each expert's feedback.

**Metamodel support for DP1 and DP2.** Regarding DP1, an area of concern raised by E2 was that he believed "determining licensing issues of legacies should be made more visible in the metamodel as it can turn out to be a major task in the migration process". In the cloud, multiple instances of a system, i.e. a virtual machine, might be created by a server based on the increased workload or rules that are triggered to scale up resources. This may cause an unintended violation of the licensing agreement that has been made between the system owner and user. This concern raised by E2 has been partially covered by concept *analyse migration cost* in the metamodel, but we have not explicated it as an individual concept in the metamodel. Utilising the knowledge source prepared during our metamodel development, a new concept named *resolve licensing issues* was added to the design phase of the metamodel (Figure 2). It is defined as follow: "Define and monitor a pay-as-you-go licensing model to handle unintended license agreement violations due to automatic scaling".

E3 explained that the metamodel lacks a concept called roll-back: "I have observed that migration process model should contain a concept to show rollback for the migration process". To address this concern, we added the concept *define roll-back plan* as a subclass of *define plan* along with a new relationship in the metamodel (Figure 2). A definition for this concept regarding the knowledge source was decided as "Define roll-back, as a plan B, to an in-house version of the legacy system in the case of occurrence of any significant risk or new application fails during the migration process. This reduces the risk and exposure to the business". With respect to DP2, there was no major comment made by the experts. E2 suggested that relationships could be added into definitions of the concepts.

**Metamodel support for DP3.** The experts provided some comments related to DP3, i.e. comprehensibility of the metamodel. From E2's viewpoint, the metamodel visualisation was unclear: "UML is not used by all stakeholders". Likewise, E4 mentioned "a unified high-level block diagram for the reference model (unifying all those three different phases) must be presented for better illustration or reflection of the model". As a response to the above comment, we had made a preliminary version of the metamodel using simple block diagrams. We believed if the metamodel is going to be an integral part of the model-driven development and OMG metamodelling framework (Atkinson & Kuhne 2003), a semi-formal representation of the metamodel becomes important when the migration scale is large. In this spirit, UML is a de-facto standard for the conceptual representation in terms of organising concepts, their relationships, and decidable reasoning.

# Discussion

## Implications for research

Existing cloud migration models do not sufficiently elaborate on process components of a legacy system migration (Fahmideh, Daneshgar, Low, & Beydoun, 2016; Jamshidi, Ahmad, & Pahl, 2013). Researchers have rather attempted to develop abstract models for cloud computing technology from different perspectives such as architectural perspective (M. Hamdaqa and Tahvildari, 2012; A. Zimmermann, et al., 2013), green cloud computing requirements (Procaccianti et al, 2014), quality aspects of cloud services (Nunez et al, 2013), code refactoring and simplification Ardagna et al. (2012), and risk and compliance effort reduction (Keller and König, 2014; Martens and Teuteberg, 2011). This paper aims at alleviating problems afflicting cloud migration from a process perspective. Our metamodel supersedes existing models in a multitude of ways.

Firstly, since the emergence of cloud computing technology, a plethora of shallow to informal models/methods have been introduced and communicated in different forms such as manual, research articles, white papers, and consulting (Fahmideh, Daneshgar, Low, & Beydoun, 2016). The models have different foci and concepts. This plethora of models brings benefits, accommodating the various ways on performing cloud migration. This raises difficulties for developers to access and to choose from, forcing developers to learn how the models work. The suggested metamodel of the current study advances our understanding about the cloud migration process and abstracts away from details and provides a platform-independent and unified view.

Secondly, the proposed metamodel provides a separation between the method design and the way the method is operationalised. Operationalisation is often bounded by underlying target cloud platforms. Separation from operationalisation issues reduces the design complexity and prevents developers from early engagement into a specific platform. This allows later them to explore each process concept in more depth. This has clear potential to improve the reusability, modularity, and maintainability of migration methods.

Finally, this research embarks on adopting a method engineering approach paving the way for hybridisation between models. Existing migration models such as such as Menychtas et al. (2013), Mohagheghi et al. (2010), Frey & Hasselbring (2011), Menzel & Ranjan (2012), Chauhan & Babar (2012), Strauch et al. (2014), and Conway & Curry (2013) largely assume that the cloud migration process is monomorphic but none is actually is a silver bullet. Each model defines a collection of activities to carry out migration process with a different scope. For instance, a model might be a better fit for moving a process-intensive and distributed workload from legacy data centers to public IaaS whilst another model maybe an adequate option to integrate legacy systems with SaaS platforms. Our metamodel provides a platform of domain concepts and their relationships grounded in what cloud computing community widely agreed. They can be selected and combined together to create a specific instance of the metamodel that fits requirements of a cloud migration project at hand. Since all of this enables a pluralist view, yet customisable and extensible, the metamodel is viewed as a configurable process model, rather than a specific model/method.

## Implications for practice

Pertaining to practice, the proposed metamodel is important in two ways. Firstly, an organisation may have its own company-wide in-house method to standardize system development but still has some deficiencies with respect to support of cloud computing concepts. The proposed metamodel can be used to augment the capability of in-house methods to carry out cloud migration projects.

Moreover, method engineers may need to select from a collection of methods that fit requirements of a given cloud migration scenario. According to Siau & Rossi (1998), metamodels are one effective way to compare family-related methods as they take place at a higher level of abstraction and capture information about methods. In the context of the cloud computing field, the proposed metamodel can act as an evaluation framework for identifying strengths, shortcomings, similarities, and differences of in-house methods. Although implied in prior studies Fahmideh et al. (2016d), this need had not been formally addressed.

Finally, from a project management point of view, the metamodel concepts can also provide an estimation of cost and heuristics of required effort to make a legacy system cloud-enabled. Our metamodel actually responds to repeated calls by studies such as Tran et al. (2011) and Quang Hieu & Asal (2012) proposing cost estimation models based on reengineering activities to be performed.

## Limitations of the study

In this paper, the metamodel applicability has been illustrated in three idiosyncratic case studies with different characteristics. However, completely satisfying the design principles can still be subject to some arguments. This is a limitation of our research as the metamodel has been only capable in representing migration models that we examined during the case analysis as a part of our metamodelling endeavour. Appraising the metamodel in different scales of cloud migration scenarios (i.e. partial or full), and organisation size (.i.e. small start-up, medium-sized organisation, and big organisations) may suggest further refinements to the metamodel concepts.

Retrospective studies present some inherent limitations (Hess, 2004). Examining the metamodel adherence to the design principles during case study analysis has relied on the accuracy of projects' databases consisting documentations, process deliverables, diagrams, and interview notes. There is a possibility of missing new concepts due to subjectivity and bias in recorded data in databases by developers. As such, there might be some concepts that could have been added to the metamodel and thus revisiting the metamodel expressivity. To alleviate this issue, we conducted follow-up communications with interviewees to confirm the validity of the secondary documents of the case studies and to provide any missing information.

With this said, although some of the refinements to the metamodel have been based on the opinions from four selected domain experts, they might have been biased and confined by their own experience and knowledge in relation to the cloud migration. Receiving feedback from a larger number of experts will reduce this threat.

## Conclusion and future research

This study was justified by the lack of a domain language for consistently representing, sharing, and standardising knowledge related to migration to the cloud catering to specific scenarios. In addressing this gap, a generic and tunable metamodel that constitutes a reusable set of domain concepts pertinent to the cloud was proposed. We have demonstrated the suitability of the metamodel in three different case studies along with positive feedback from domain experts.

The current study points to few directions for further research. The metamodel augmentation with new concepts relevant to the post-migration, for example, continuous integration and delivery, is a possible area of improvement. Similarly, the metamodel can be extended by incorporating concepts related to the growing area of mobile cloud applications that are ran on mobile devices (Dinh, Lee, Niyato, & Wang, 2013; Giurgiu, Riva, Juric, Krivulev, & Alonso, 2009). Such applications are characterised with challenges such as battery life, bandwidth, heterogeneity, and privacy that arise in mobile environments. The UML formalism used in the proposed metamodel representation can facilitate the inclusion new further concepts in a structured way.

The metamodel instantiation for the creation of situational methods involves some factors such as the choice of a target cloud platform, the pricing model of cloud providers and characteristics of the development team. Making trade-offs among these factors that sometimes contradict or depend on each other has an impact on the metamodel instantiation and specialization. In future work, one can utilise the idea of goal-driven method tailoring suggested in (Cesar & Paolo, 2009; Karlsson & Ågerfalk, 2011) as a baseline to define mechanisms for selecting metamodel concepts and putting them together to create method instances for a particular cloud migration scenario. We expect that this research will motivate other researchers to further explore new approaches which will systematically facilitate cloud computing adoption.